\documentclass{elsart1p}

\usepackage{graphicx}

\usepackage{amsmath}
\usepackage{amssymb}
\usepackage{cite}
\usepackage{tabularx}

\usepackage{dcolumn}
\usepackage{bm}

\begin{document}

\begin{frontmatter}


\title{Analytic continuation in two-color QCD: new results on the critical line}

\author[label1]{P. Cea}
\author[label2]{L. Cosmai}
\author[label3]{M. D'Elia}
\author[label4]{A.Papa}
\address[label1]{Dipartimento di Fisica Univ. Bari \& INFN - Bari - Italy}
\address[label2]{INFN - Bari - Italy}
\address[label3]{Dipartimento di Fisica Univ. Genova \& INFN - Genova - Italy}
\address[label4]{Dipartimento di Fisica Univ. Calabria  \&INFN - Cosenza - Italy}

\begin{abstract}
We test the method of analytic continuation from imaginary to real chemical potential
in two-color QCD, which is free from the sign problem.
In particular, we consider the analytic continuation of the critical
line to real  values of the  chemical potential.
\end{abstract}

\end{frontmatter}

\section{Introduction}
\label{Introduction}
Understanding the phase diagram of QCD on the temperature-chemical potential ($T,\mu$) has many important implications in cosmology, in astrophysics and in the phenomenology 
of heavy ion collisions.
It is known that the fermion determinant in QCD becomes complex in presence of a non-zero chemical potential, thus preventing us from performing standard Monte Carlo simulations. 
Various techniques have been exploited
to circumvent this problem.
Among these techniques we consider here the  use of an imaginary chemical potential
for analytic
continuation~\cite{deForcrand:2002ci,deForcrand:2003hx,D'Elia:2002gd,D'Elia:2004at,Azcoiti:2005tv,Chen:2004tb,Giudice:2004se,Kim:2005ck,Cea:2006yd,Papa:2006jv,deForcrand:2006pv,deForcrand:2007rq,D'Elia:2007ke,Conradi:2007be,Cea:2007wa,Karbstein:2006er}.
It is very important to have some control on the method of analytic continuation
and a direct test of this method in simpler models can be very useful.
To this purpose we explored the case of 2-color QCD where, for any value of the chemical potential, the fermion matrix is real. 
In ref.~\cite{Cea:2006yd} we found, by looking at the behavior in both real and imaginary chemical potential of the Polyakov loop, the chiral condensate and the fermion density, that the method gives reliable results, within appropriate ranges of the chemical potential, and that a considerable improvement can be achieved if ratio of polynomials are used to interpolate data with imaginary chemical potential~\cite{Cea:2006yd,Papa:2006jv} .

On the other hand one of the most important applications of analytic continuation is
the determination of the critical line or of critical surfaces
for small values of $\mu$~\cite{deForcrand:2002ci,deForcrand:2003hx,D'Elia:2002gd,D'Elia:2004at,Azcoiti:2005tv,Chen:2004tb,deForcrand:2006pv,deForcrand:2007rq}.
The theoretical basis in this case is
not as straightforward as for physical observables and relies on
the assumption that susceptibilities, whose peaks signal the presence
of the transition, be analytic functions of the parameters on a finite
volume~\cite{deForcrand:2002ci,deForcrand:2003hx}. Direct tests of the method are
even more important in this case, therefore we
extended~\cite{Cea:2007vt} our analysis of two-color QCD to the study of the
(pseudo)critical line.
As in usual QCD simulations,
we will determine locations of the critical line for
$\mu^2 < 0$ and interpolate them by
suitable functions to be continued to $\mu^2 > 0$. The prediction
obtained at real $\mu$ will then be compared with direct
determinations of the transition line.

\begin{figure}[t!]
\vspace{-0.05cm}
\begin{center}
\includegraphics*[width=0.70\columnwidth]{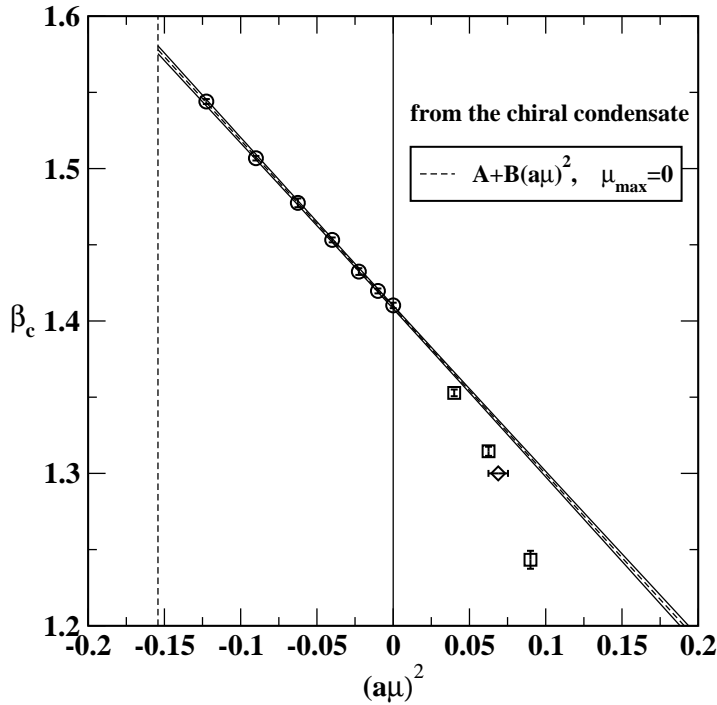}
\end{center}
\vspace{-0.3cm}
\caption{
Critical couplings obtained from the chiral susceptibility, together
with a
linear fit in $(a\mu)^2$ to data with $\mu^2 \leq 0$.
The datum at
$\beta_c=1.30$ (diamond) is taken from Ref.~\cite{Conradi:2007be}.
}
\label{fig:crit_psibpsi}
\vspace{-0.2cm}
\end{figure}

\begin{figure}[t!]
\vspace{-0.05cm}
\begin{center}
\includegraphics*[width=0.70\columnwidth]{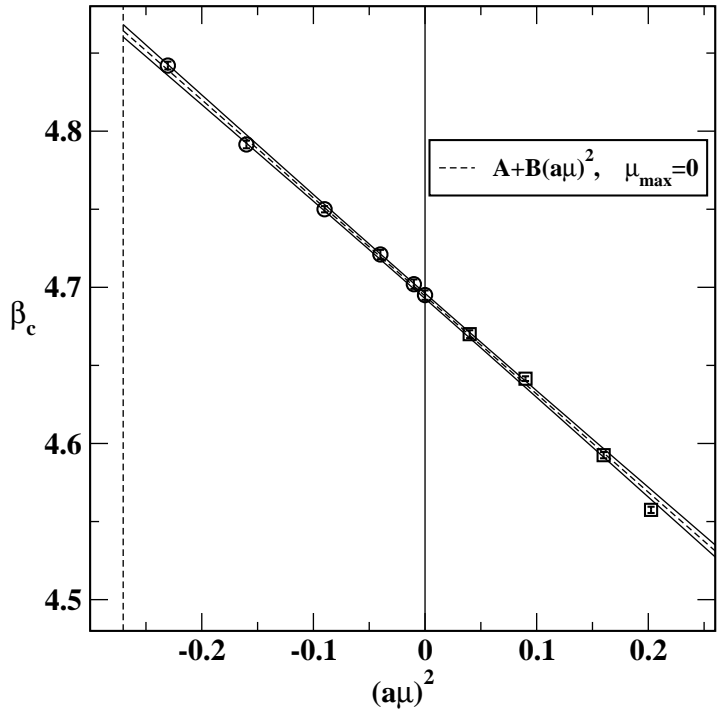}
\end{center}
\vspace{-0.3cm}
\caption{
Our preliminary results for the critical line in SU(3) at finite isospin on a $8^3\times4$ lattice 
and $am=0.1$, together
with a
linear fit in $(a\mu)^2$ to data with $\mu^2 \leq 0$.
}
\label{fig:finisospin}
\vspace{-0.2cm}
\end{figure}

\section{Lattice simulations}
\label{Lattice}
Numerical simulations have been performed using hybrid Monte Carlo 
with $dt=0.01$ . Typical statistics collected corresponds to   ~ 20k trajectories.
Simulations have been done using resources of the INFN apeNEXT computing center. 

Our aims are finding the best interterpolation of critical line at imaginary chemical potential and
testing the analytic continuation of the critical line from imaginary to real chemical potential.
We considered  SU(2) gauge theory with $N_f=8$ staggered fermions with mass $a m = 0.07$
on a $16^3 \times 4$ lattice~\cite{Cea:2007vt}.
In Fig.~1 we present results for the extrapolation of the critical line to $\mu^2>0$ compared with
the direct determination of the critical coupling $\beta_c$. As one can see there is a discrepancy between the extrapolated critical line and the direct determination of $\beta_c(\mu^2)$.  One can infer that or the critical line is not analytic in the whole range of $\mu^2$ considered here or that the interpolation at $\mu^2 \leq 0$ is not accurate enough to correctly reproduce the behavior at $\mu^2 > 0$.
A common fit to all available data with a third order polynomial
in $\mu2$ suggests the second possibility may be the correct one.
In order to go deep inside this question  we have started new simulations using a different mass ($am =0.2$) on a $16^3 \times 4$ lattice and mass $a m = 0.07$ on a $16^3 \times 6$ lattice. 
We have also considered the test of the  analytical continuation of the critical line the case
of SU(3) with finite isospin and $N_f=8$ staggered fermions with mass $a m = 0.1$ on a 
$8^3 \times 4$ lattice.
Fig.~2 displays our preliminary results for the analytic continuation of the critical line 
in the case of SU(3) at finite isospin density. As one can see the discrepancy between the extrapolated critical line and the direct determination of $\beta_c(\mu^2)$ is less severe. 

\section{Conclusions}
\label{Conclusions}
We have tested in two-color QCD the analytic continuation of the critical line in the 
$(T,\mu)$  plane from imaginary to real chemical potential.
We have found that the critical line around  $\mu=0$ can  be described by an analytic function. Indeed, a third order polynomial in $\mu^2$ fits all the available data for the critical coupling.
However, when trying to infer the behavior of the critical line at real $\mu$   from the extrapolation of its behavior at imaginary $\mu$, a very large precision would be needed to get the correct result.
In the case of polynomial interpolations there is a clear indication that high-order terms play a relevant role at $\mu^2>0$  but are less visible at $\mu^2<0$, this calling for an accurate knowledge of the critical line in all the first Roberge-Weiss sector.  This scenario could be peculiar of two-color QCD. 
If confirmed in other theories free of the sign problem, such as QCD at finite isospin density, then one should seriously reconsider the analytic continuation of the critical line in the physically relevant case of QCD at finite baryon density.

For the reason above we have started  a systematic program of investigations both in two-color QCD and in SU(3) QCD at finite isospin.
Our preliminary results suggest that it may be useful to revisit the interpolations used in QCD (first order polynomial in $\mu^2$) and  make the effort of determining accurately at least one more term in the polynomial fit.










\end{document}